# Mobsimilarity: Vector Graph Optimization for Mobility Tableau Comparison

Yuhao Yao, Haoran Zhang, Jinyu Chen, Wenjing Li, Mariko Shibasaki, Ryosuke Shibasaki, *Member, IEEE* and Xuan Song

**Abstract**—Human mobility similarity comparison plays a critical role in mobility estimation/prediction model evaluation, mobility clustering and mobility matching, which exerts an enormous impact on improving urban mobility, accessibility, and reliability. By expanding origin-destination matrix, we propose a concept named mobility tableau, which is an aggregated tableau representation to the population flow distributed between different location pairs of a study site and can be represented by a vector graph. Compared with traditional OD matrix-based mobility comparison, mobility tableau comparison provides high-dimensional similarity information, including volume similarity, spatial similarity, mass inclusiveness and structure similarity. A novel mobility tableaus similarity measurement method is proposed by optimizing the least spatial cost of transforming the vector graph for one mobility tableau into the other and is optimized to be efficient. The robustness of the measure is supported through several sensitive analysis on GPS based mobility tableau. The better performance of the approach compared with traditional mobility comparison methods in two case studies demonstrate the practicality and superiority, while one study is estimated mobility tableaus validation and the other is different cities' mobility tableaus comparison.

## 1 INTRODUCTION

Modeling dynamic human mobility provides sufficient information for urban planning [1], transportation management [2-4], ride sharing system [5], emergency management [6] and prevention of infectious diseases [7], which exerts an enormous impact on improving urban mobility, accessibility, and reliability.

Human mobility similarity comparison plays a critical role in modeling dynamic human mobility. Basically, mobility similarity is the key indicator to evaluate the quality of mobility observing datasets. A variety of mobility observing datasets has been widely utilized to extract large-scale population flow information (including GPS [8], Bluetooth [9], GSM [10], automatic number plate recognition (ANPR) sensors [11, 12] and smartphone Call Detailed Record (CDR) [13, 14]). Due to the different observing mechanisms, gaps exist in the qualities of these data sources. Mobility similarity comparison is a high-dimensional and homogeneous method to evaluate these gaps; Additionally, mobility similarity comparison can be naturally used for mobility clustering, which distills general rules from messy mobility information [15] and mines the coupling relationship of urban subspace and dynamic causes [16]; Moreover, mobility similarity comparison also helps to match the mobility between different regions from a policymaker perspective. For first-tier cities, there are usually abundant data of human mobility and full-fledged data-driven application such as public transportation management [17], traffic congestion avoiding plan [18], car sharing system [19], etc. However, such human mobility data collection and data-driven services are much less in small cities. A well-designed mobility similarity measurement can provide abundant useful information for transferring mature applications from one city into another by mobility features.

Origin-destination (OD) matrix is an effective and the most widely used form to describe the high-dimensional information of human mobility. OD matrix emphasizes on the volume of population flow between different region pair and exerts an enormous function on numerous fields. However, people flow usually consists of numerous basic patterns, including commuting, working and commercial. Such a complex spatial-temporal data is extremely difficult to extract useful information directly by the volume similarity provided by OD matrix, thus requires high dimensional abstract features to describe them [20]. The research gap remains in effectively describing the high-dimensional information.

To bridge the gaps, in this paper, we proposed a concept named mobility tableau, which is a geographical correlation related aggregated tableau representation to the population flow distributed between different location pairs. Additionally, we developed a corresponding method for mobility tableau comparison.

The organization of the remaining sections is: In section 2, we summarize the related works and the current limitations. The motivation and framework of this new comparison method are shown in section 3. Section 4 formulates the mobility tableau. The corresponding method for mobility tableau comparison is illustrated in section 5. Section 6 and 7 display the results of the numerical experiments and two real applications respectively. Conclusions are given in the last section.

• *Y. Yao, H. Zhang, J. Chen, W. Li, M. Shibasaki, R. Shibasaki and X. Song are with the Center for Spatial Information Science, the University of Tokyo, Tokyo, Japan. E-mail: (yaoyh, zhang_ronan, miraclec, liwenjing, shiba, songxuan)@csis.u-tokyo.ac.jp, kk106816ms@gmail.com.*

## 2 RELATED WORKS

In literatures, numerous traditional metrics such as RMSE [21-23], MSE [24], GU [25], E [26], $r^2$ [27] can well solve the volume similarity measurement problem based on OD matrices. However, method that target on multi-dimensional similarity measurement for mobile tableau is sparse. Djukic et al. [28] first put forward a Mean structural similarity index (MSSIM). Based on this work, MSSIM variants such as a 4D-MSSIM is proposed by Van Vuren et al. [29], geographical window based structural similarity index (GSSI) is proposed by Behara et al. [30] and a spatially weighted structural similarity index (SpSSIM) is proposed by Jin et al. [31]. Ruiz de Villa et al. [32] propose a Wasserstein distance based method and Behara et al. [33] propose a Levenshtein distance based method.

In this section, we divide mobility tableau similarity comparison mathods in literatures into three types based on their key ideas: volume difference focused measure, image-based measure and transforming distance-based measure.

### 2.1 Volume Difference Focused Measure

The key idea among volume difference focused measures is to regard population flow of different origin-destination pair as an independent statistic, then traditional statistical metrics can be applied to measure the volume (dis)similarity of two mobility tableau.

Here several notable measures are listed: root mean square error (RMSE) [21-23]; normalised root mean square error (RMSN) [34]; mean square error (MSE) [24]; mean absolute error ratio (MAER) [35]; mean absolute percent error (MAPE) [36]; goodness of Theil's fit (GU) [25]; maximum possible absolute error (MPAE) [37]; relative error (RE) [38]; total demand deviation (TDD) [39]; R-squared ($R^2$) [27], and entropy measure (E) [26].

Although aforementioned measures have simple mathematical formulations and compute deviations between individual population flows, they can just simply reflect the volume similarity of mobility tableaus and ignore all other properties.

### 2.2 Image-based Measure

#### *2.2.1 Mean structural SIMilarity index (MSSIM)*

MSSIM [40] is the prototype to compare the structural degradation between two images by comparing pixels. Since cells of matrix are similar with pixels in an image, Djukic et al. [28] applied MSSIM on OD matrices.

Three different components – luminance $l$, contrast $c$, and structure $s$ are introduced in $SSIM$, which can be calculated by mean, standard deviation and covariance.

#### *2.2.2 MSSIM's Variants*

Based on MSSIM, several improvements have been made in the literature:

- In 4D-MSSIM proposed by Van Vuren and Day-Pollard [29], neighborhood OD pairs are identified using spatial proximity between OD pairs by using Euclidian distance.
- In GSSI proposed by Behara et al. [30], geographical boundaries of higher-level zones are used to define the local windows;
- In SpSSIM proposed by Jin et al. [31], a series of spatial weight matrices are used to take place the original moving window.

Although the problem about the sensitivity of MSSIM towards local window size and geographical adjacency of zones have been addressed, some problems still remain, especially issues related to stability constants and network specific parameters selection.

### 2.3 Transforming Distance-based Measure

The main idea of distance-based measure is to measure mobility similarity by computing the least cost of transforming one mobility tableau into the other.

#### *2.3.1 Wasserstein Metric*

One of the transforming distance-based measure utilizes Wasserstein distance (Ruiz de Villa et al. [32]), which used to be widely applied in mass transportation problems such as optimal cost required to transfer iron-ore from many mining locations to several factories etc.

In the context of mobility tableaus, the Wasserstein distance between two mobility tableaus is defined as the minimum total travel time required to assign the trips between origin-destination pairs of mobility tableau $X$ and $Y$. By creating virtual population flow to boundary, $X$ and $Y$ can have same total population flows.

Although this metric can measure the structural similarity by taking geographic correlation into account, it has two fatal limitations: First, as an optimization problem, when the number of regions is too large, it is computationally very expensive; second, when the total population flow number of two mobility tableau is different, there is no definite way to balance the difference, so the result is not unique.

#### *2.3.2 Levenshtein Metric*

Another measure selects Levenstein distance (Behara et al. [33]), which used to be a measure of proximity between two strings. To be specific, Levenstein distance calculates least expensive set of insertions, deletions or substitutions that are required to transform one string into another.

Each origin's preferred destination and corresponding demand value forms a sequence as string, which can be transformed into the other. By computing the cost of the transforming, LODn as an absolute comparison of $n$th rows can be achieved. By normalizing LODn, a relative comparison with the trip productions (sum of origin flows) for $n$th row from both matrices as NLODn can be get.

The overall comparison between the mobility tableaus is obtained through mean Levenshtein distance named LOD, which is the average of all LODn values, and the mean normalised Levenshtein distance named NLOD, which is the average of all NLODn.

This metric clearly distinguishes the structural similarity from the volume similarity of mobility tableaus. However, the geographic correlation of different OD pairs is ignored, which will cause the result worse when mobility tableaus have high sptial similarity.

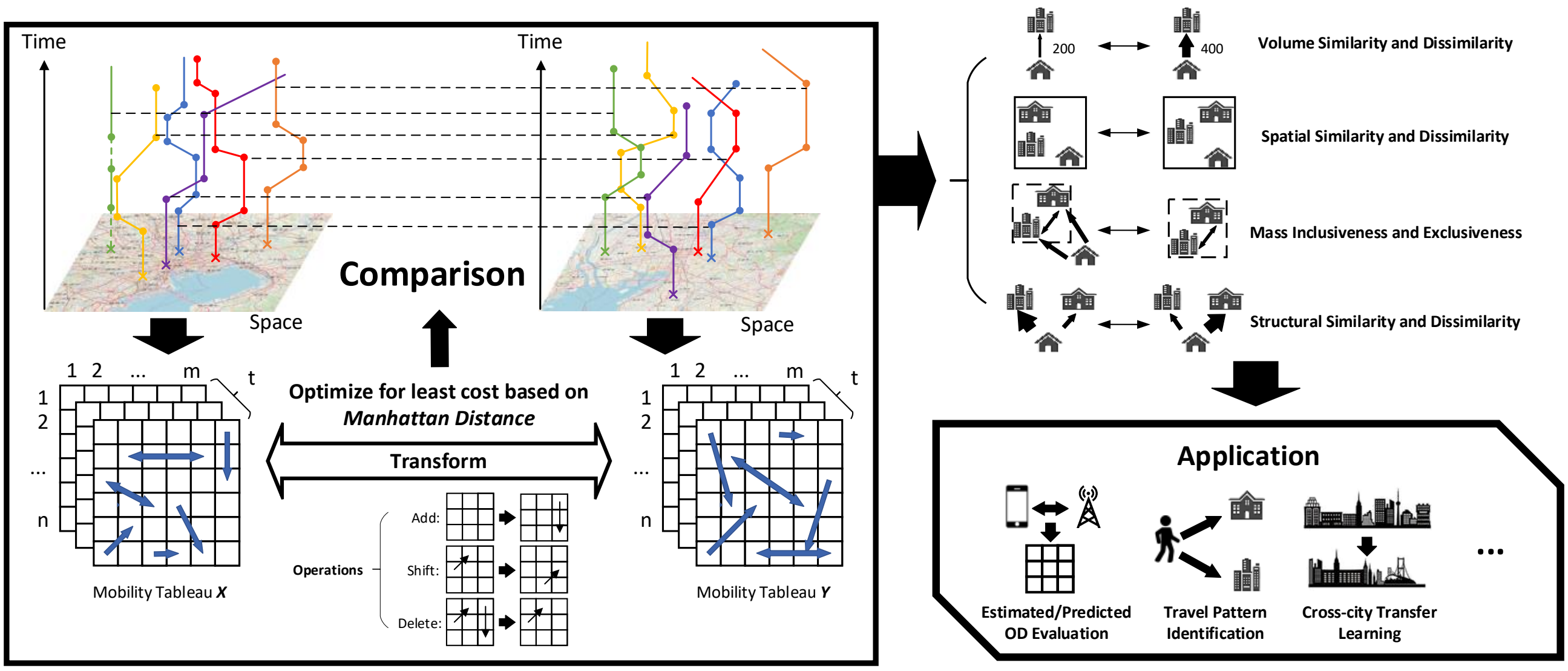

Fig. 1. Framework of vector graph optimization for mobility tableau comparison

### 2.4 Contributions of this work

To summarise, even though all these studies can reflect the volume similarity of population flow, very limited studies are able to reflect the structural similarity. Further, studies that focus on spatial similarity or geographical correlation related structural similarity are even less. On the other hand, several structural focused methods need to set proper parameters, which is not suitable for all cases.

Compared with these studies, our method has several contributions:

- We propose a metric for mobility tableau comparison, which provides high-dimensional similarity information, including volume similarity, spatial similarity, mass inclusiveness and structure similarity.
- We present the theoretical formulation of mobility tableau similarity and mathematically prove it workable.
- We propose the computation approach for mobility tableau similarity and optimize it to be efficient.
- Our approach is applied to two real world case studies and the better performance of our approach compared with traditional mobility comparison method demonstrate our practicality and superiority.

## 3 Framework

To further explore mobility similarity measurement problem, we need to introduce a new indicator which can be an aggregated representation to the population flow distributed between different location pairs of a study site. Compared with traditional OD matrix comparison, the new indicator should be geographic correlation related, so the comparison could not only reflect the volume similarity of each population flow as OD matrix, but also demonstrate several significant physical properties, including spatial similarity, mass inclusiveness and structural similarity.

To this end, we propose a novel mobility similarity measure by computing the least spatial cost of transforming the vector graph for one mobility tableau into the other. To be specific, three basic operations and corresponding cost based on Manhattan distance are defined for vector graph transforming. The least cost of transforming can be computed by an optimization model and further simplified into a bipartite graph maximum weight matching problem solved by a point-oriented Kuhn-Munkres algorithm. Several metrics are defined to well explain the (dis)similarity between two mobility tableaus in different respects. The framework is shown as Fig. 1.

The robustness of the measure is tested through several sensitive analysis on GPS based mobility tableau.

By the proposed method, numerous practical applications can be achieved, such as mobility tableau estimation/prediction model performance evaluation, Travel pattern identification, cross-city transfer learning etc. Two real world application case studies help to demonstrate the practicality of the approach while one is estimated mobility tableaus validation and the other is different cities' mobility tableaus comparison.

## 4 Mobility Tableau Formulation

A mobility tableau can be described by a vector graph, in which the grid represents the map of the study cite while vectors represent the volume of population flow and the corresponding origin/destination regions. Compared with traditional OD matrix comparison, mobility tableau is geographic correlation related, so the comparison between two mobility tableaus can not only reflect the volume similarity of each population flow as OD matrix, but also demonstrate several significant physical properties, including spatial similarity, mass inclusiveness and structural similarity.

To illustrate mobility tableau and three important properties of mobility tableau similarity, we demonstrate three cases that shown in Fig. 2. Assume that all study sites are 3×3 grid, and each cell represents a region. There is always an omitted regular population flow between any two regions. By using arrow as vector to represent huge population flow, a vector graph can be constructed to describe the mobility tableau.

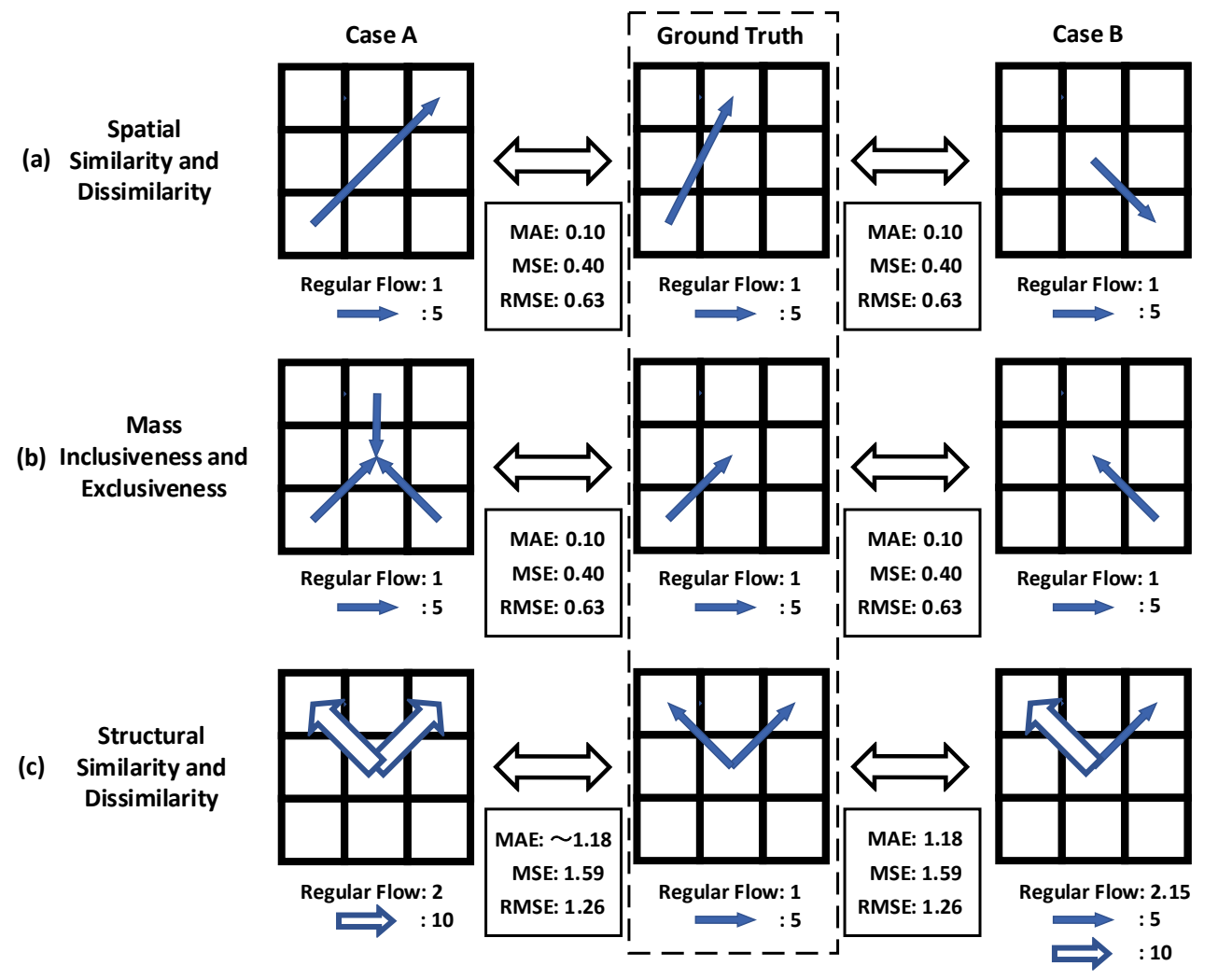

Fig. 2. Demonstration of three physical properties of mobility similarity

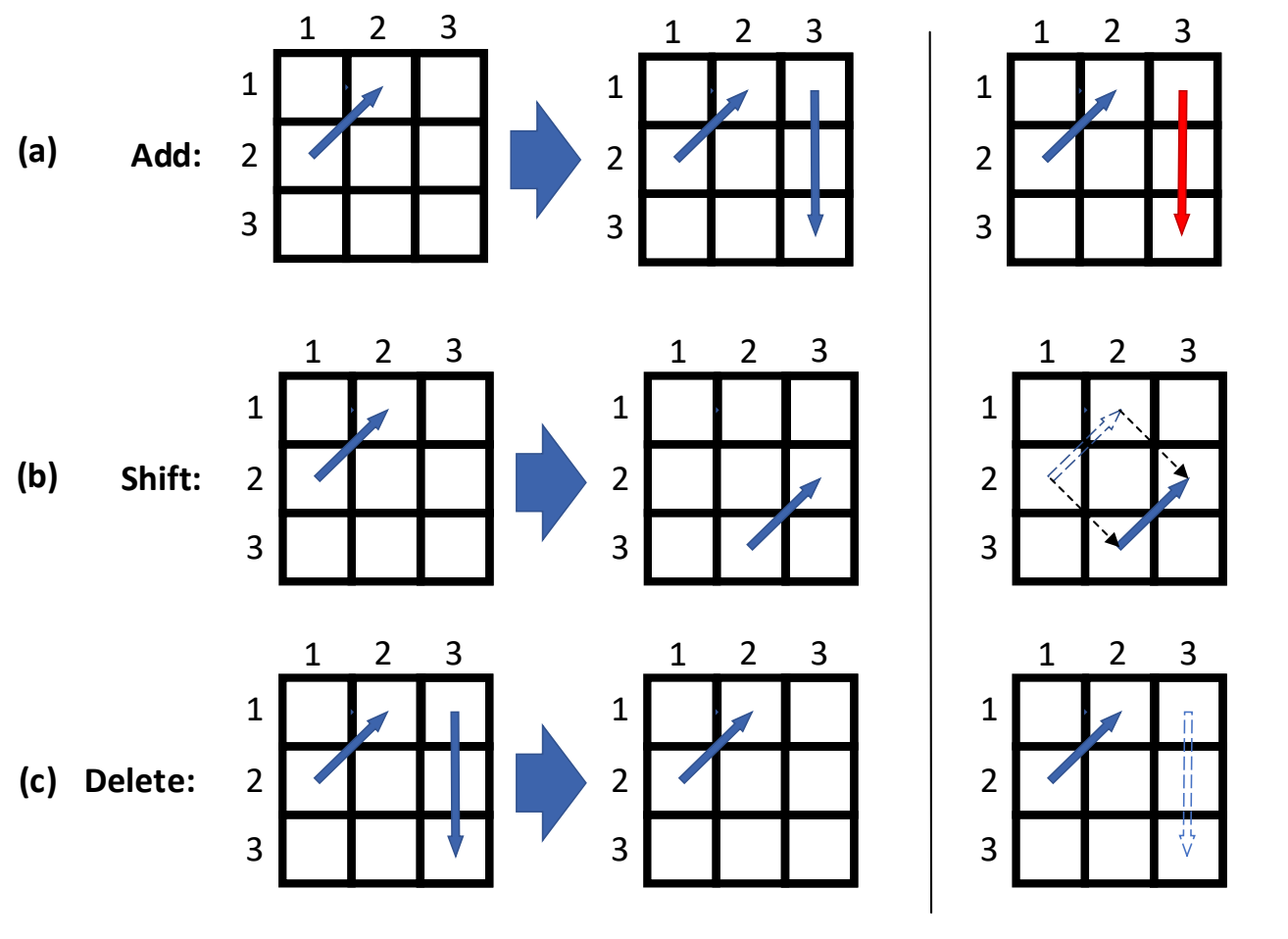

Fig. 3. Basic operations for vector graph transformation. In case (a), (b) and (c), the cost of operation equals to $\|0 \rightarrow \vec{\delta}_{(3,1),(3,3)}\| = |3-3| + |3-1| = 2$ , $\|\vec{\delta}_{(1,2),(2,1)} \rightarrow \vec{\delta}_{(2,3),(3,2)}\| = |2-1| + |3-2| + |3-2| + |2-1| = 4$ and $\|\vec{\delta}_{(3,1),(3,3)} \rightarrow 0\| = |3-3| + |3-1| = 2$, relatively.

Spatial similarity and dissimilarity represent whether two mobility tableaus are similar by taking geographic correlation into account. In Fig. 2. (a), though compared with ground truth all traditional metrics are the same for case A and case B, without no doubt case A is much closer to ground truth because the only huge population flow is closer in space. Therefore, case A has much more spatial similarity to ground truth than case B.

Mass inclusiveness and exclusiveness represent the extent for one mobility tableau to cover the other. In Fig. 2. (b), though compared with ground truth all traditional metrics are the same for case A and case B, three huge population flows in case A totally cover the only huge population flow in ground truth, while in case B the huge population flow is exclusive from the ground truth. Therefore, case A has much more mass inclusiveness to ground truth than case B.

Structural similarity and dissimilarity represent the (dis)similarity of the proportion of population flows from the same origin region to different regions between two mobility tableaus. In Fig. 2. (c), though compared with ground truth all traditional metrics are similar for case A and case B, case A is just double the population flow of ground truth, which indicates they have the same structure, while the structure of case B is totally different. Therefore, case A has much more structural similarity to ground truth than case B.

If we divide the geographic location of the study site $M$ into $m \times n$ mesh-grids based on longitudes and latitudes range, the volume of population flow for an origin-destination pair starts with cell $(ox, oy)$ and ends with cell $(dx, dy)$ in the mobility tableau $X$ will be $X_{(ox,oy),(dx,dy)}$, $ox, dx \in [1, m], oy, dy \in [1, n]$.

We represent this OD pair by a vector $\vec{\delta}_{(ox,oy),(dx,dy)}$, here vector $\vec{\delta}$ is not a free vector so the start point and the end point are important characteristics. By using "$*$" to represent the number of vectors, the mobility tableau $X$ can be represented by a vector graph (or vector set) $S_X$ constructed by these vectors as:

$$S_X = \left\{X_{(1,1),(1,1)} * \vec{\delta}_{(1,1),(1,1)}, \dots, X_{(oi,oj),(di,dj)} * \vec{\delta}_{(oi,oj),(di,dj)}, \dots, X_{(m,n),(m,n)} * \vec{\delta}_{(m,n),(m,n)}\right\} \quad (1)$$

Measuring the similarity between mobility tableaus $X$ and $Y$ is equivalent to measuring the similarity between the vector graph $S_X$ and $S_Y$.

## 5 Mobility Tableau Comparison

### 5.1 Proposed Transforming Operation

To further explain how to measure the similarity between two different OD pair sets, we define an operator "$\rightarrow$" for vector or vector set named *Transformation*. $S_X \rightarrow S_Y$ means the transformation from vector graph $S_X$ to vector graph $S_Y$.

To conduct the transformation, three basic operations are defined as Fig. 3. shows: $0 \rightarrow \vec{\delta}$ named *Add*, $\vec{\delta} \rightarrow \vec{\delta}'$ named *Shift* and $\vec{\delta} \rightarrow 0$ named *Delete*.

- **Add:** Add represents directly adding a new vector into the graph. For example, as Fig. 3. (a) shows, when $S_X = \{\vec{\delta}_{(1,2),(2,1)}\}, S_Y = \{\vec{\delta}_{(1,2),(2,1)}, \vec{\delta}_{(3,1),(3,3)}\}$, to conduct $S_X \rightarrow S_Y$, the red vector $\vec{\delta}_{(3,1),(3,3)}$ is added.
- **Shift:** Shift represents shifting a vector from its original position to another, which changes the vector from one into another. For example, as Fig. 3. (b) shows, when $S_X = \{\vec{\delta}_{(1,2),(2,1)}\}, S_Y = \{\vec{\delta}_{(2,3),(3,2)}\}$ , to conduct $S_X \rightarrow S_Y$ , the vector $\vec{\delta}_{(1,2),(2,1)}$ is shifted from $(1,2) \rightarrow (2,1)$ to $(2,3) \rightarrow (3,2)$.
- **Delete:** Delete represents directly deleting a vector from the set. For example, as Fig. 3. (c) shows, when $S_X = \{\vec{\delta}_{(1,2),(2,1)}, \vec{\delta}_{(3,1),(3,3)}\}, S_Y = \{\vec{\delta}_{(1,2),(2,1)}\}$, to conduct $S_X \rightarrow S_Y$, the vector $\vec{\delta}_{(3,1),(3,3)}$ is deleted.

By basic operations, transformation between any two vector graphs becomes possible. The similarity between two vector graphs can be indicated by the least cost of the transformation between them. We select Manhattan Distance as cost function and define an operator "‖ ‖" to represent the cost of each operation. The cost of basic operations is listed below:

- **Add:** For add operation, the cost equals to the Manhattan distance between added vector's start point and end point, i.e.

$$\left\|0 \to \vec{\delta}_{(ox,oy),(dx,dy)}\right\| = |dx - ox| + |dy - oy| \quad (2)$$

(13)

- **Shift:** In Shift operation, two shift vectors are generated between the previous start point and new start point, and previous end point and new end point. The cost of shift operation equals to the sum of these two vectors' Manhattan distance between start point and end point, i.e.

$$\left\|\vec{\delta}_{(ox,oy),(dx,dy)} \to \vec{\delta}_{(ox',oy'),(dx',dy')}\right\| = |ox' - ox| + |oy' - oy| + |dx' - dx| + |dy' - dy| \quad (3)$$

- **Delete:** For delete operation, the cost is similar with add operation, which equals to the Manhattan distance between deleted vector's start point and end point, i.e.

$$\left\|\vec{\delta}_{(ox,oy),(dx,dy)} \to 0\right\| = |dx - ox| + |dy - oy| \quad (4)$$

(15)

We define a function $V(S_X \to S_Y)$ as the least cost of a transformation $S_X \to S_Y$. Because Manhattan distance has a property named trigonometric inequality, which indicates that the direct Manhattan distance from point A to another point B will be no more than the Manhattan distance passing a third point C, for any vector $\vec{\delta}$ and $\vec{\delta}'$ we have:

$$V(\emptyset \to \{\vec{\delta}\}) = \|0 \to \vec{\delta}\| \le \|0 \to \vec{\delta}'\| + \|\vec{\delta}' \to \vec{\delta}\| \quad (5)$$

Then we can easily get:

$$V(S_X \to \emptyset) = \sum \|\vec{\delta}_i \to 0\|, \vec{\delta}_i \in S_X \quad (6)$$

$$V(\emptyset \to S_Y) = \sum \|0 \to \vec{\delta}'_i\|, \vec{\delta}'_i \in S_X' \quad (7)$$

$$V\left(\{\vec{\delta}\} \to \{\vec{\delta'}\}\right) = min\,(\|\vec{\delta} \to \vec{\delta'}\|, \|\vec{\delta} \to 0\| + \|0 \to \vec{\delta'}\|) \quad (8)$$

The least cost $V(S_X \to S_Y)$ can be computed iteratively:

$$V(S_X \to S_Y) = min\Big(V(S_X' \to S_Y') + V\big((S_X - S_X') \to (S_Y - S_Y')\big), V(S_X' \to \emptyset) + V\big((S_X - S_X') \to S_Y\big), V(\emptyset \to S_Y') + V(S_X \to (S_Y - S_Y'))\Big), \forall S_X' \subseteq S_X, \forall S_Y' \subseteq S_Y \quad (9)$$

## 5.2 Operational Properties of Cost Function

Cost function $V$ has several operational properties that can be proved, which helps to reduce the complexity of computation.

$$V(S_X \to S_Y) = V(S_Y \to S_X) \quad (10)$$

### *5.2.1 Reversibility of Cost Function*

Because for any vector $\vec{\delta}$ and $\vec{\delta}'$ we have $\|\vec{\delta} \to 0\|$=$\|0 \to \vec{\delta}\|$ and $\|\vec{\delta} \to \vec{\delta}'\| = \|\vec{\delta}' \to \vec{\delta}\|$, the reversibility of transformation between any two vector sets can be simply proved. The reversibility guarantees that the transformation cost between two mobility tableaus has no relationship with the transformation direction, so it is possible to select the source vector set which takes less time for computation.

### *5.2.2 Trigonometric Inequality of Cost Function*

$$V(S_X \to S_Y) \le V(S_X \to S') + V(S' \to S_Y) \quad (11)$$

Here $S'$ represents any vector set including empty set. This property is extended from Manhattan distance's trigonometric inequality. This property makes sure that there will be no other vectors generated during the transforming process. Therefore, OD pair that does not have population flow do not need to be considered in computation.

### *5.2.3 Reducibility of Cost Function*

For any $V(S_X \to S_Y)$, there must exist vector sets $A \subseteq S_X - S'$ and $B \subseteq S_Y - S'$ that for $\forall S' \subseteq S_X \cap S_Y$, $V(S_X \to S_Y) = V\big((S_X - S' - A) \to (S_Y - S' - B)\big) + V(A \to S') + V(S' \to B)$.

$$V(S_X \to S_Y) = V\big((S_X - S') \to (S_Y - S')\big), \forall S' \subseteq S_X \cap S_Y \quad (12)$$

Because of (11) we have $V(A \to S') + V(S' \to B) \ge V(A \to B)$, so $V(S_X \to S_Y) \ge V\big((S_X - S' - A) \to (S_Y - S' - B)\big) + V(A \to B)$. Because of (9), we have $V\big((S_X - S' - A) \to (S_Y - S' - B)\big) + V(A \to B) \ge V\big((S_X - S') \to (S_Y - S')\big)$, so $V(S_X \to S_Y) \ge V\big((S_X - S') \to (S_Y - S')\big)$. Because of (9), we also have $V(S_X \to S_Y) = V((S_X - S' + S') \to (S_Y - S' + S')) \le V\big((S_X - S') \to (S_Y - S')\big) + V(S' \to S') = V\big((S_X - S') \to (S_Y - S')\big)$. Therefore, $V(S_X \to S_Y) = V\big((S_X - S') \to (S_Y - S')\big)$.

This property dramatically simplifies the computation because it proves for any set $S_X$ and $S_Y$, we have $V(S_X \to S_Y) = V((S_X - S_X \cap S_Y) \to (S_Y - S_X \cap S_Y))$, which significantly reduces the number of vectors.

## 5.3 Physical Properties Characterization

### 5.3.1 Volume Similarity

Obviously, the least cost $V(S_X \to S_Y)$ reflects the volume difference between two vector graphs $S_X$ and $S_Y$ of mobility tableaus by taking geographical correlation into account. Considering about $V(S_X \to S_Y)$ is affected by cell width $w$, we compute Manhattan distance for mobility tableaus (MD) by:

$$MD(X,Y) = V(S_X \to S_Y) * w \quad (13)$$

To measure the mass similarity, we normalize mass difference by computing the cost from empty set to $S_X$ and $S_Y$ to get normalized Manhattan distance for mobility tableaus (NMD) by:

$$NMD(X,Y) = 1 - \frac{V(S_X \to S_Y)}{V(\emptyset \to S_X) + V(\emptyset \to S_Y)} \quad (14)$$

Because of the reversibility and trigonometric inequality of cost function that introduced in section 5.2.2, we have $V(\emptyset \to S_X) + V(\emptyset \to S_Y) = V(S_X \to \emptyset) + V(\emptyset \to S_Y) \geq V(S_X \to S_Y)$, so NMD varies from 0 to 1, while the bigger NMD is, the more similar two mobility tableaus are. Specifically, when NMD=1, two mobility tableaus are exact the same one; when NMD=0, two mobility tableaus have no similarity at all.

### 5.3.2 Spatial Similarity

For a transformation $\{\vec{\delta}\} \to \{\vec{\delta}'\}$, there will be two possible implementation ways: $\vec{\delta} \to \vec{\delta}'$ and $(\vec{\delta} \to 0) + (0 \to \vec{\delta}')$. Assume that $\vec{\delta}$ is the vector from point $A$ to point $B$ while $\vec{\delta}'$ is the vector from point $A'$ to point $B'$. Measuring the cost of two ways is equivalent to comparing the Manhattan length between two pairs of opposite sides of the quadrilateral formed by $ABA'B'$, i.e. $AB + A'B'$ and $AA' + BB'$. When $AB + A'B'$ is longer than $AA' + BB'$, the transformation prefers *Shift* operation because $\delta$ and $\delta'$ are close in space. On the other hand, when $AA' + BB'$ is longer, the transformation prefers *Add* and *Delete* operation because $\vec{\delta}$ and $\vec{\delta}'$ are quite far away. The total cost of *Shift* operation and *Add/Delete* operation just correspondingly indicate that whether the error comes from Spatial Similarity and Spatial Dissimilarity.

Therefore, if we decompose function $V$ into $V_{add/delete}$ and $V_{shift}$ where $V_{add/delete}$ is the total cost of *Add/Delete* operation and $V_{shift}$ is the total cost of *Shift* operation, a shift proportion (SP) can be computed as:

$$SP(X,Y) = \frac{V_{shift}(S_X \to S_Y)}{V_{shift}(S_X \to S_Y) + V_{add/delete}(S_X \to S_Y)} \quad (15)$$

$SP$ illustrates the proportion of mobility tableaus difference that can be corrected by spatial similarity.

### 5.3.3 Mass Inclusiveness

For vector sets $S_X$ and $S_Y$, if we put them into a Euclidean space, in which the distance between two vector set equals to the least cost of Manhattan distance between them and use point $A$, $B$ to represent them correspondingly, a triangle can be formed by introducing an empty set as point $O$ as Fig. 4. shows. Obiviously the angle of $\angle AOB$ can reflect the direction difference between $S_X$ and $S_Y$. Here direction refers to the process to build the mobility tableau. Same direction means building one mobility tableau needs the same process of building the other one while contrary direction means there is no common process between building two mobility tableaus.

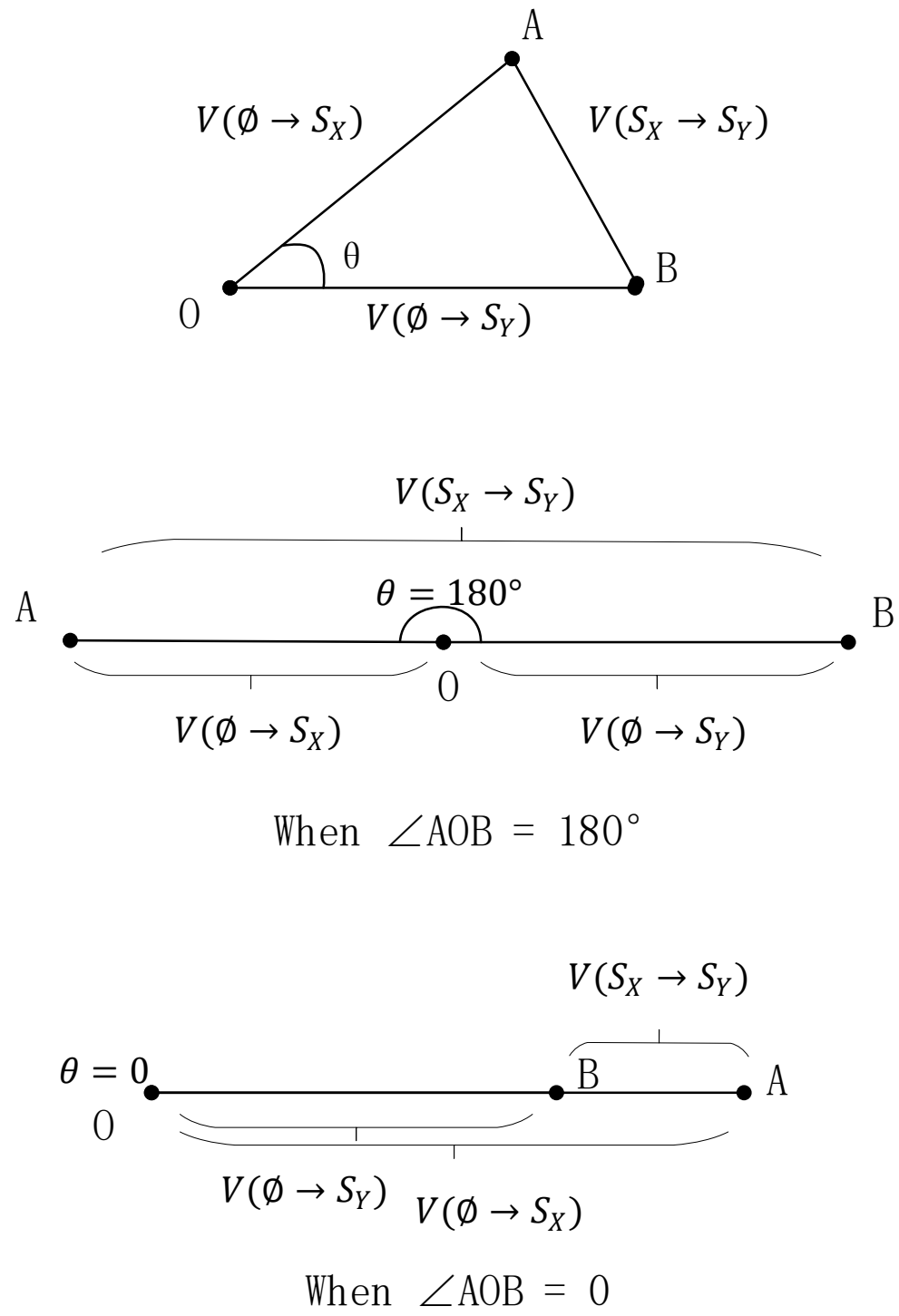

Fig. 4. Manhattan distance triangle for two mobility tableaus

We define a normalized mass angle (NMA) which is related to $\angle AOB$ as:

$$NMA = 1 - \frac{arccos\frac{V(\emptyset \to S_X)^2 + V(\emptyset \to S_Y)^2 - V(S_X \to S_Y)^2}{2V(\emptyset \to S_X) * V(\emptyset \to S_Y)}}{\pi} \quad (16)$$

Specifically, when NMA=1 i.e. $\angle AOB = 0$, one mobility tableau totally cover the other, i.e. they have inclusiveness; when NMA=0 i.e. $\angle AOB = 180°$, in order to transform into the other, one mobility tableau need to delete itself, which means two mobility tableaus are totally in the different directions and have exclusiveness.

### 5.3.4 Structural Similarity

In section 5.3.3, when $\theta = 0$, one mobility tableau may cover the other instead of equaling to the other because the total number of OD flow is different. Therefore, the structural (dis)similarity metric can be computed in the same way by normalizing two mobility tableaus.

Assume that $X'$ and $Y'$ are normalized mobility tableaus for $X$ and $Y$. Then normalized structural angle (NSA) can be defined similar with (17) by:

$$NSA = 1 - \frac{arccos\frac{V(\emptyset \to S_{X'})^2 + V(\emptyset \to S_{Y'})^2 - V(S_{X'} \to S_{Y'})^2}{2V(\emptyset \to S_{X'}) * V(\emptyset \to S_{Y'})}}{\pi} \quad (17)$$

Specifically, when $NSA = 1$, two mobility tableaus have the same structure; when $NSA = 0$, in order to transform into the other, one normalized mobility tableau need to delete itself, which means two normalized mobility tableaus are totally in the different directions, i.e. they have no structural similarity.

However, in the real case, $NSA = 0$ will almost never happen because when the total population flow of mobility tableau is large, two mobility tableaus seldom have no spatial similarity. Therefore, NSA will be usually very high. Considering about two random mobility tableaus will also have a relatively stable NSA $\mu$, we define a randomness removed normalized structural angle (RRNSA) as:

$$RRNSA = \begin{cases} \frac{NSA-\mu}{1-\mu}, & \frac{NSA-\mu}{1-\mu} > 0 \\ 0, & \frac{NSA-\mu}{1-\mu} \leq 0 \end{cases} \quad (18)$$

$\mu$ will be determined in section 6.1.1.

## 5.4 Method of Computation

### 5.4.1 Optimization Model

For the given mobility tableaus $X$ and $Y$ in the study site $R$, convert them into vector sets $S_X$ and $S_Y$, then compute the origin vector set $S_X' = S_X - S_X \cap S_Y$, and destination vector set $S_Y' = S_Y - S_X \cap S_Y$. The least cost can be computed by a linear optimization model, of which the objective function is as follow:

$$min\ (\sum_i(\alpha_i * D_i) + \sum_j(\beta_j * A_j) + \sum_{i,j}(\gamma_{i,j} * T_{i,j})) \quad (19)$$

Subject to:

$$M_i = D_i + \sum_j T_{i,j}, \forall i \in I \quad (20)$$
$$N_j = A_j + \sum_i T_{i,j}, \forall j \in J \quad (21)$$

Notations are listed in Table 1. However, assume that the total number of different areas in $R$ is $k$, in the worst case, there are $1/2k^2$ types of vector both in origin vector set and destination vector set, and therefore variables will be more than $1/4k^4$, which will cost too much time on computation.

### 5.4.2 Point Set oriented Kuhn-Munkres Algorithm

By constructing a bipartite graph network, the problem of computing the least cost of transforming one vector set into another can be converted into finding the edge set with maximum weight, of which all vertices are included and just included once.

The most popular algorithm to solve this problem in a polynomial time is the Kuhn-Munkres algorithm, which was proposed by Kuhn [41]. Munkres observed that Kuhn-Munkres algorithm is strongly polynomial [42]. Edmonds and Karp [43], and independently Tomizawa modified the algorithm to achieve an O(n[3]) running time. Theoretically, the Kuhn-Munkres algorithm is guaranteed to reach the global optimal. More information can be found refers to [44].

TABLE 1
NOTATIONS OF OPTIMIZATION MODEL

| Notation | Description |
|---|---|
| $\boldsymbol{I}$ | The total number of vector type in origin vector set. |
| $\boldsymbol{J}$ | The total number of vector type in destination vector set. |
| $\boldsymbol{M_i}$ | The number of the $i$-th type of vector in origin vector set, $i \in I$. |
| $\boldsymbol{N_j}$ | The number of the $j$-th type of vector in destination vector set, $j \in J$. |
| $\boldsymbol{D_i}$ | The number of the $i$-th type of vector in origin vector set that is deleted by delete operation, $i \in I$. |
| $\boldsymbol{A_j}$ | The number of the $j$-th type of vector in destination vector set that is generated by add operation, $j \in J$. |
| $\boldsymbol{T_{i,j}}$ | The number of the $i$-th type of vector in origin vector set that is transformed into the $j$-th type of vector in destination vector set by shift operation, $i \in I, j \in J$. |
| $\boldsymbol{\alpha_i}$ | The Manhattan length of the $i$-th type of vector in origin vector set, $i \in I$. |
| $\boldsymbol{\beta_j}$ | The Manhattan length of the $j$-th type of vector in destination vector set, $j \in J$. |
| $\boldsymbol{\gamma_{i,j}}$ | The Manhattan distance of the $i$-th type of vector in origin vector set transforming into the $j$-th type of vector in destination vector set, $i \in I, j \in J$. |

For our case, we construct a bipartite graph network $G(V, E)$, where $V = V_X \cup V_Y$. In $V$, each type of vector in $S_X'$ is a vertex in $V_X$ and each type of vector in $S_Y'$ is a vertex in $V_Y$, without no doubt we have $V_X \cap V_Y = \emptyset$. Additional labels $p(u)$ is initialized as the number of vectors for each vertex (vector type) to save the number of vectors for each vertex (vector type) that has not found a match. Construct edge set by $E = V_X \times V_Y$, the weight of edge $w$ is computed by:

$$w(i, j) = -\gamma_{i,j}, \forall i \in V_M, \forall j \in V_N \quad (22)$$

Introduce one zero vector into $V_X$ and $V_Y$ respectively as $v_{YO}$ and $v_{XO}$. Update edge set $E$, corresponding weight $w$ and $p$ as:

$$w(v_{XO}, v_{YO}) = 0 \quad (23)$$
$$w(i, v_{XO}) = -\alpha_i, \forall i \in (V_X - v_{XO}) \quad (24)$$
$$w(v_{YO}, j) = -\beta_j, \forall j \in (V_Y - v_{YO}) \quad (25)$$
$$p(v_{XO}) = \sum_{i \in V_Y} N_i \quad (26)$$
$$p(v_{YO}) = \sum_{i \in V_X} M_i \quad (27)$$

We use a set $F_G(S)$ to represent vertex set $S$'s neighborhood, which includes all the vertices that share an edge with vertices in $S$. Define labels $l(u)$ for each vertex in graph, with each label of a vertex corresponding to its only matching constraint. A dictionary $D(i), i \in V_N$ is used to store vectors' distribution for each vertex $i$. From $D(i)$ we can get the matching $R_D$. For a matching $R_D$ and vertex $v$, if we have $p(v) = 0$, we call the vector matched, otherwise we call it free. We denote by $G_l$ the subgraph of $G$ which contains those edges where $l(i) + l(j) = w(i, j)$. $G_l$ is a spanning subgraph of $G$, and includes all vertices from $G$ but only those edges from the bipartite matching, which allows the vertices to be perfectly feasible. Then the pseudo code of a point set-oriented Kuhn-Munkres algorithm can be described as Table 2.

TABLE 2
POINT SET ORIENTED KUHN-MUNKRES ALGORITHM

**Algorithm** Point Set oriented Kuhn-Munkres algorithm

**Input:** A bipartite graph $\boldsymbol{G} = (\boldsymbol{V_X} \cup \boldsymbol{V_Y}, \boldsymbol{E})$, weight $\boldsymbol{w(i,j)}$, different vertex's number $\boldsymbol{p(u)}$.

**Output:** Optimal perfect matching $\boldsymbol{R_D}$.

**Step 1:** Generate initial labelling $\boldsymbol{l}$, $\boldsymbol{D}$ and matching $\boldsymbol{R_D}$ in $\boldsymbol{G_l}$.

$$\boldsymbol{l(i) = \max(w(i,j), \forall j \in V_Y), \forall i \in V_X}$$
$$\boldsymbol{l(j) = 0, \forall j \in V_Y}$$

**Step 2:** If $\boldsymbol{R_D}$ perfect, stop. Otherwise pick free vertex $\boldsymbol{i \in V_X}$. Set $\boldsymbol{S = i, T = \emptyset}$.

**Step 3:** If $\boldsymbol{F_G(S) = T}$, update labels (forcing $\boldsymbol{F_G(S) \neq T}$):

$$\boldsymbol{g_l = \min_{s \in S, j \notin T}\{l(i) + l(j) - w(i,j)\}}$$
$$\boldsymbol{\hat{l}(u) = \begin{cases} l(u) - g_l, i \in S \\ l(u) + g_l, i \in T \\ l(u), otherwise \end{cases}}$$

**Step 3:** If $\boldsymbol{F_G(S) \neq T}$, choose $\boldsymbol{j \in F_G(S) - T}$:

- If $\boldsymbol{j}$ free and $\boldsymbol{p(i) > p(j)}$. Set $\boldsymbol{p(i) = p(i) - p(j)}$, $\boldsymbol{p(j) = 0}$, $\boldsymbol{T = T \cup j}$. Go to Step 3.
- If $\boldsymbol{j}$ free and $\boldsymbol{p(i) = p(j)}$. Set $\boldsymbol{p(i) = 0}$, $\boldsymbol{p(j) = 0}$, $\boldsymbol{T = T \cup j}$. Go to Step 2.
- If $\boldsymbol{j}$ free and $\boldsymbol{p(i) < p(j)}$. Set $\boldsymbol{p(i) = 0}$, $\boldsymbol{p(j) = p(j) - p(i)}$. Go to Step 3.
- If $\boldsymbol{j}$ is matched, set $\boldsymbol{T = T \cup j}$. For each $\boldsymbol{k \in D(j)}$, extend alternating tree: $\boldsymbol{S = S \cup k}$. Go to Step 3.

## 6 SENSITIVE ANALYSIS

In this section, we determine parameter $\mu$ mentioned in 5.3.3 and test the robustness of all the metrics proposed in this paper.

**Study site and dataset:** Tokyo, Japan is selected as the study site and the dataset is a part of "Konzatu-Tokei(R)" Data provided by Zenrin DataCom INC. "Konzatsu-Tokei (R)" Data refers to people flows data collected by individual location data sent from mobile phone under users' consent, through Applications provided by NTT DOCOMO, INC. Those data is processed collectively and statistically in order to conceal the private information. Original location data is GPS data (latitude, longitude) sent in about every a minimum period of 5 minutes and does not include the information to specify individual. ※Some applications such as "docomomap navi" service (map navi・local guide).

The reference mobility tableau is a GPS data-based vector graph in a 40 × 40 grid (with 1 km width cell), observed from January 1st, 2011 to January 31st, 2011, which involves more than 2,500 individuals.

One experiment is set for parameter determining and four experiments are set to test different indicators, which is described as Table 3.

### 6.1 Experimental Set up

#### 6.1.1 Parameter Determining

In this test, we search for the average NSA between two irrelevant mobility tableaus as parameter $\mu$. To achieve it, we first construct a mobility tableau $X$, then we shuffle all regions to get a new mobility tableau $Y$. Calculate NSA between $X$ and $Y$. RRNSA is robust if NSA between $X$ and $Y$ is close to a constant value $\mu$.

#### 6.1.2 Criteria for uniform scaling effects

In this test, we change the original mobility tableau with different uniform scaling percentages. The reference mobility tableau, $X$ is compared with $Y$ where $Y = \varphi * X$, $\varphi$ is chosen from [0.1, 0.2, 0.3…1.9, 2.0]. MD, NMD, RRNSA are robust metrics if:

- MD is 0 when $\varphi = 1$ and should increase as deviated from unity.
- NMD is 100% when $\varphi = 1$ and should decrease as deviated from unity.
- RRNSA is always 100%.

TABLE 3
EXPERIMENT DESCRIPTION

| | Content | Tested Metrics: MD (volume similarity) | NMD (volume similarity) | SP (sptial similarity) | NMA (mass inclusiveness) | RRNSA (structural similarity) |
|---|---|---|---|---|---|---|
| **Experiment 1** | Parameter determining | | | | | ✓ |
| **Experiment 2** | Uniform scaling effects | ✓ | ✓ | | | ✓ |
| **Experiment 3** | Random scaling effects | | | | | ✓ |
| **Experiment 4** | Random spatial exchange | | | ✓ | | |
| **Experiment 5** | Random data source change | | | | ✓ | |

### 6.1.3 Criteria for random scaling effects

In this test, we change the original mobility tableau with uniform scaling percentages adding random scaling percentages. The reference mobility tableau, $X$ is compared with $Y$ where $Y = X * (\varphi + \omega * rand[0,1])$, in which $\varphi$ is chosen from [0.6, 0.8, 1.05] and $\omega$ is chosen from [0.05, 0.1, 0.15, 0.2]. Here different $\varphi$ represents different demand scenarios that are encountered in traffic demand modeling [45]. RRNSA is a robust metric if:

- For the same $\varphi$, RRNSA decreases when $\omega$ increases.
- For the same $\omega$, RRNSA increases when $\varphi$ increases.

### 6.1.4 Criteria for spatial exchange

In this test, we change the original mobility tableau by exchanging regions in two different ways several times. We first define two different region exchanging way: Neighborhood exchange and random exchange. Neighborhood exchange means exchanging two regions which are adjacent while random exchange means randomly exchange two regions regardless of distance. For the reference mobility tableau $X$, we transform it into compared mobility tableau $Y$ by conducting region exchanging 5 times, in which there are $k$ times neighborhood exchange and $5 - k$ times random exchange, $k \in [0,1,2,3,4,5]$. SP is a robust metric if SP increases when $k$ increases.

### 6.1.5 Criteria for random data source

In this test, we change the data source of the original mobility tableau by mixing with other data source. In this case, assume that the original mobility tableau $X$ is estimated by 1,000 individuals, denoted by $U$. To estimate the compared mobility tableau $Y$, we select $1000 - k * 100$ people from $U$, and $k * 100$ people not from $U$ as the data source. Then NMA is robust if NMA decreases when $k$ increases.

## 6.2 Results of Sensitive Analysis

### 6.2.1 Parameter Determining

The result of parameter determining shows that the mean value of NSA for random mobility tableaus equals to 75.22%, while the standard deviation is about 0.01, which is relatively concentrated and proves that NSA is robust at a certain extent. We pick the mean value 75.22% of NSA for random mobility tableaus as $\mu$.

### 6.2.2 Result of Uniform Scaling Effects

The result of uniform scaling effects is shown as Fig. 5. In this graph, the unit of MD is $10^6$ km. From the figure it can be seen that:

- MD is 0 when $\varphi = 1$, which means two matrices are the same one. When $\varphi$ increases/decreases from 1, MD linearly changes corresponding to $\varphi$, which indicates that the difference between two matrices grows up.
- NMD is 100% when $\varphi = 1$, which means two matrices are the same one. When $\varphi$ increases/decreases from 1, NMD decreases, which indicates that the difference between two matrices grows up.
- RRNSA is always 100%, which means two mobility tableaus always have the same structure.

The above satisfies the robustness condition outlined in Section 6.1.2, and therefore we can conclude that MD, NMD and RRNSA is robust towards uniform scaling effects.

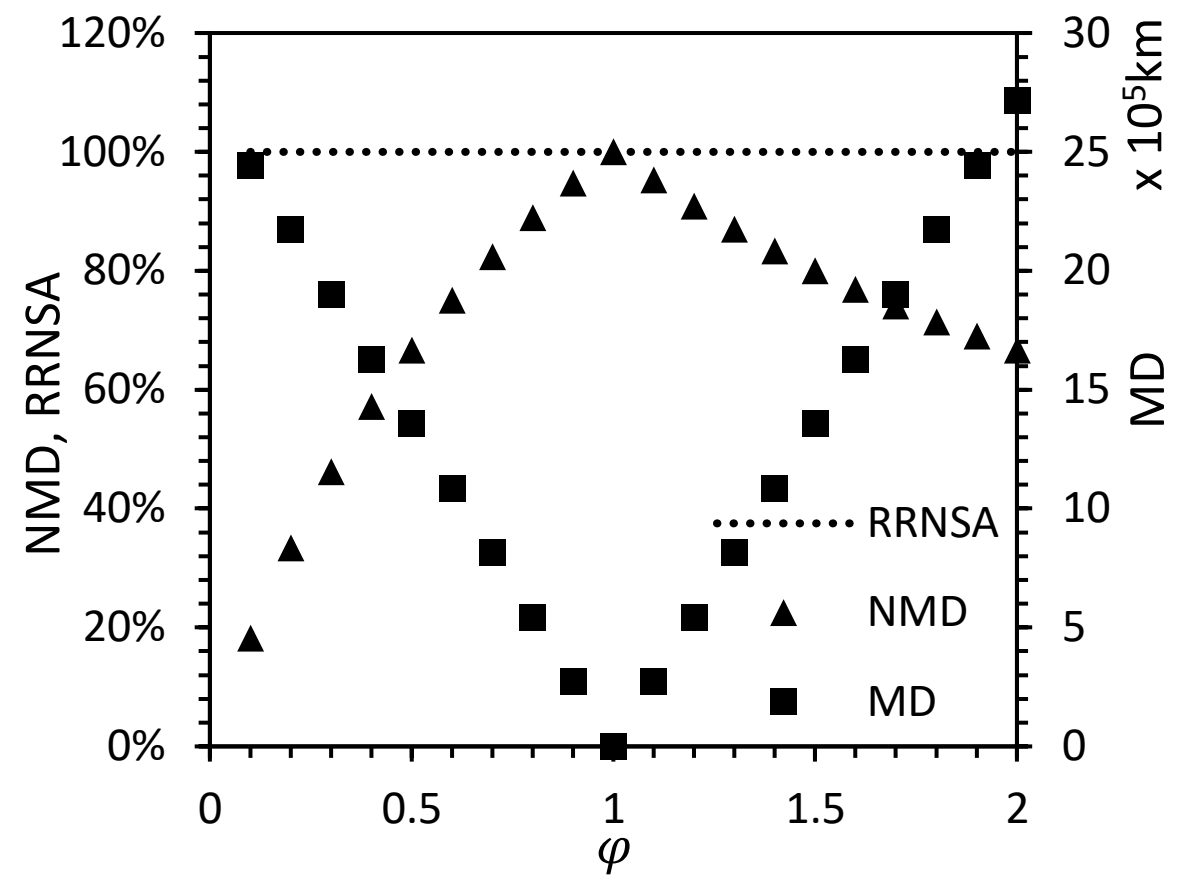

Fig. 5. NMD, RRNSA and MD with different $\varphi$ in uniform scaling effects

### 6.2.3 Result of Random Scaling Effects

The result of random scaling effects for 100 times is shown as Fig. 6. The variance of RRNSA for different $\varphi$ and $\omega$ is always smaller than $10^{-5}$, which validates that the mean value of RRNSA can represent all cases. From the figure we can see that:

- When $\varphi$ is fixed, the mean value of RRNSA decreases when $\omega$ increases, which indicates that when randomness becomes higher, the structure similarity metric RRNSA becomes lower.
- When $\omega$ is fixed, the mean value of RRNSA increases when $\varphi$ increases, which indicates that when

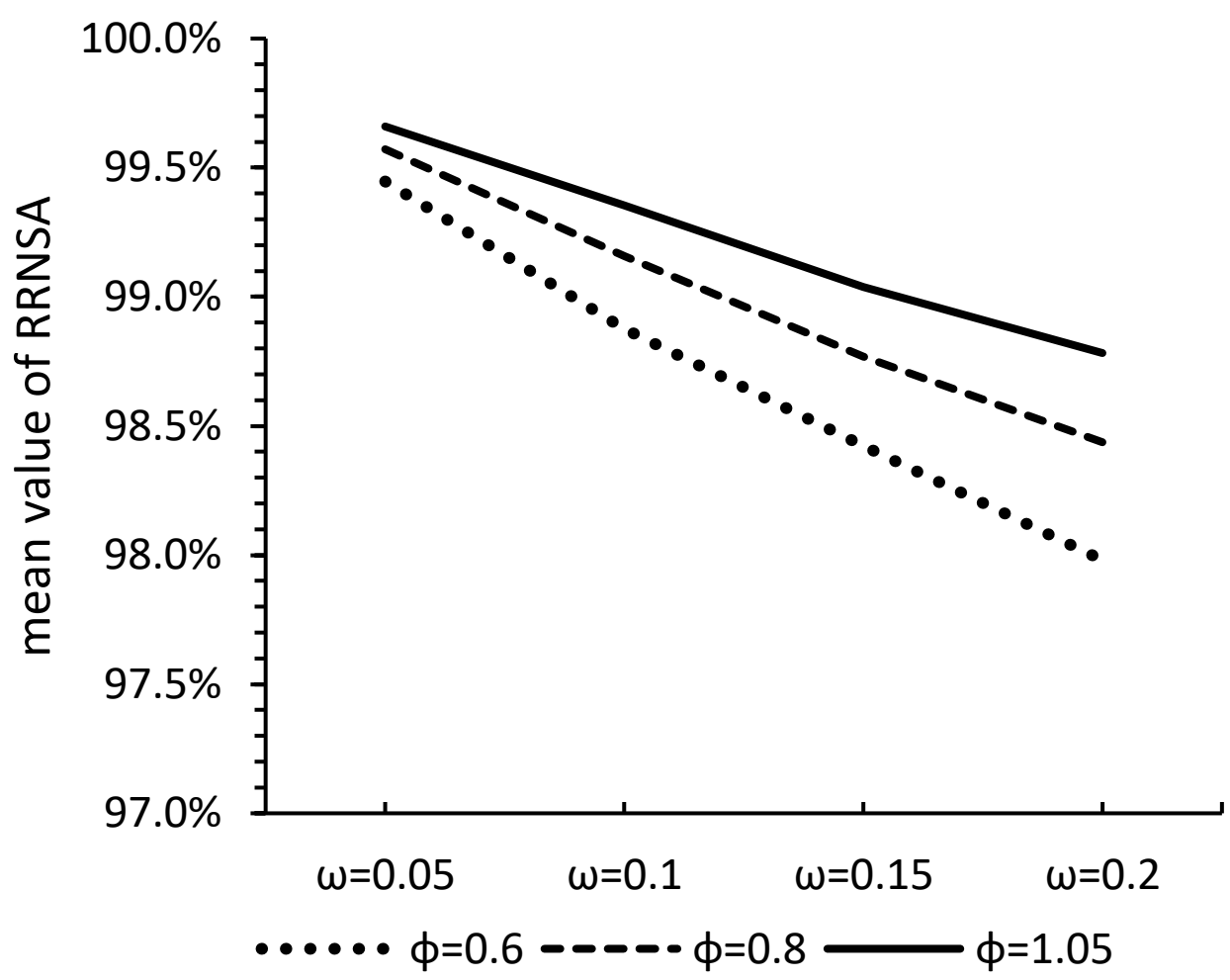

Fig. 6. Mean value of RRNSA for different $\varphi$ and $\omega$ in random scaling effects

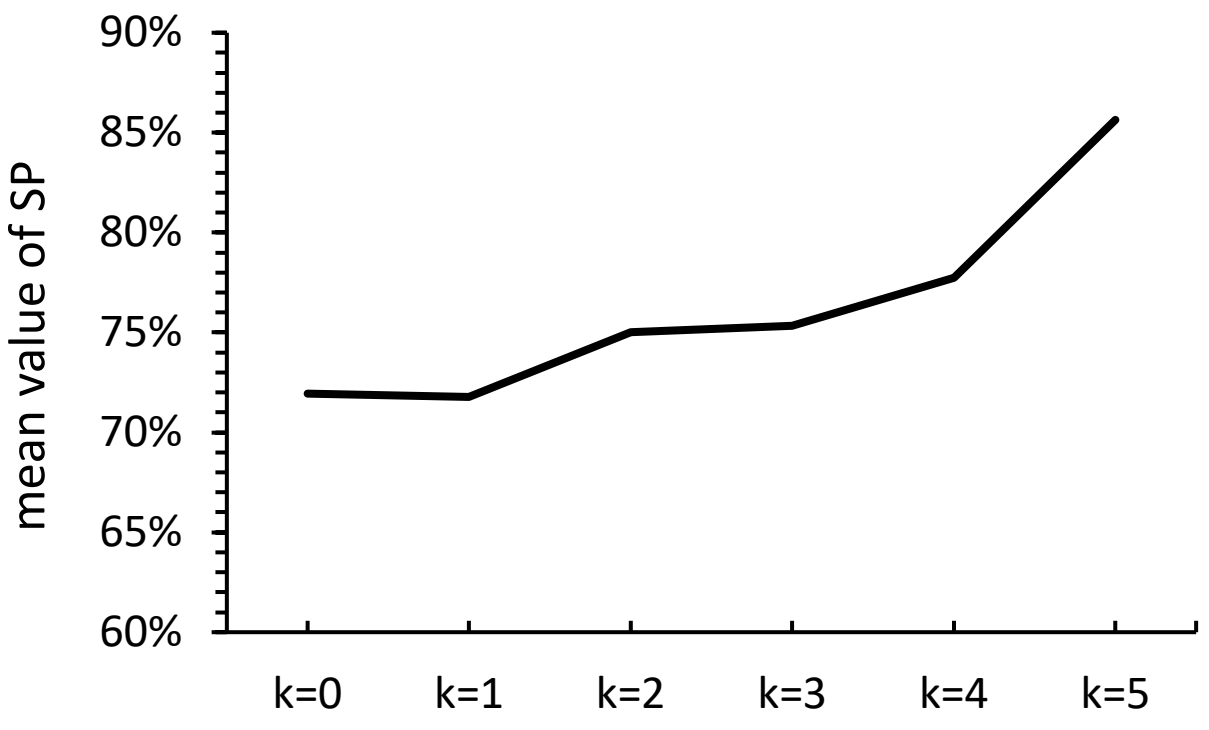

Fig. 7. Mean value of SP for different k in random spatial exchange

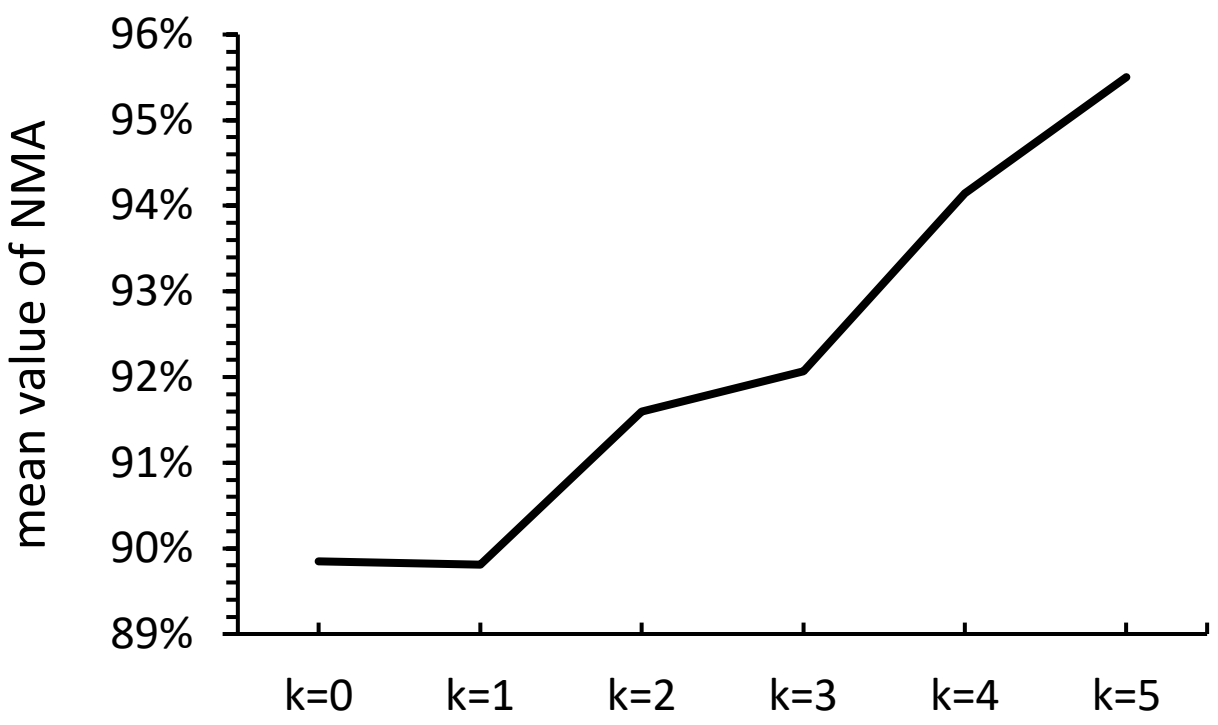

Fig. 8. Mean value of NMA for different k in random data source change

randomness becomes lower, the structure similarity metric RRNSA becomes higher.

The above satisfies the robustness condition outlined in Section 6.1.3, and therefore we can conclude that RRNSA is robust towards random scaling effects.

#### 6.2.4 *Result of Random Spatial Exchange*

The result of random spatial exchange for 100 times is shown as Fig. 7. Although because the randomness of sptial exchange is quite large, the variance of SP for different *k* is about 0.02 and the mean value of SP is not staple enough, we can still conclude it from the figure that the changing tendency of SP is becoming large when k increases, which means when two matrices become closer in spatial, the spatial similarity metric SP will be much higher.

The above satisfies the robustness condition outlined in Section 6.1.4, and therefore we can conclude that SP is robust towards random spatial exchange.

#### 6.2.5 *Result of Random Data Source Change*

The result of uniform data source change for 100 times is shown as Fig. 8. From the figure we can see that although the randomness of data source change is so large that the NMA is not very concentrated, the changing tendency of NMA is still clear that it becomes large when k increases, which means when one matrix becomes more possible to cover the other, the mass invlusiveness metric NMA will be much larger.

The above satisfies the robustness condition outlined in Section 6.1.5, and therefore we can conclude that NMA is robust towards random data source change.

## 7 Real Case Application

In this section, we demonstrate two real world application case studies to confirm the practicality of the approach.

The first one is an estimated mobility tableau validation. As we introduced before, mobility estimation is extremely important and attracts a wide range of research interest. Therefore, a proper validation method is significant to measure the quality of estimated result. We demonstrate an example of validating CDR based estimated mobility tableau by the proposed method.

The second case study is different city mobility tableau comparison. If we divide city into slices and based on mobility similarity to match slices between different cities, then mobility related application can be transferred from one city into another. Therefore, we demonstrate an example of comparing mobility tableau between two different cities to find a best match of city slices.

### 7.1 Estimated Mobility Tableau Validation

In this section, we compare the mobility tableau estimated from CDR data with the mobility tableau estimated from GPS data. CDR data is generated when the mobile device has any data interation with neighboring cell tower, it records the position of the cell tower and the time.

An example of an estimated OD trajectory from CDR data is demonstrated by Fig. 9. Since the location of CDR data is the position of the cell tower, there is usually a distance between CDR data and the real location. From the figure we can see that structures of two trajectories are very similar, but once if we focus on the grid based estimated OD pair, we will find it very different from the ground truth. This phenomenon indicates that we cannot simply measure the similarity between two mobility patterns by the volume similarity of corresponding OD matrices.

Compared with CDR records, GPS records have high accuracy that are closed to ground truth. Therefore, in this case study we regard GPS based estimated mobility tableau as ground truth to validate CDR based estimated mobility tableau. Two mobility tableaus are both from the same population, which includes 2,500 individuals in one

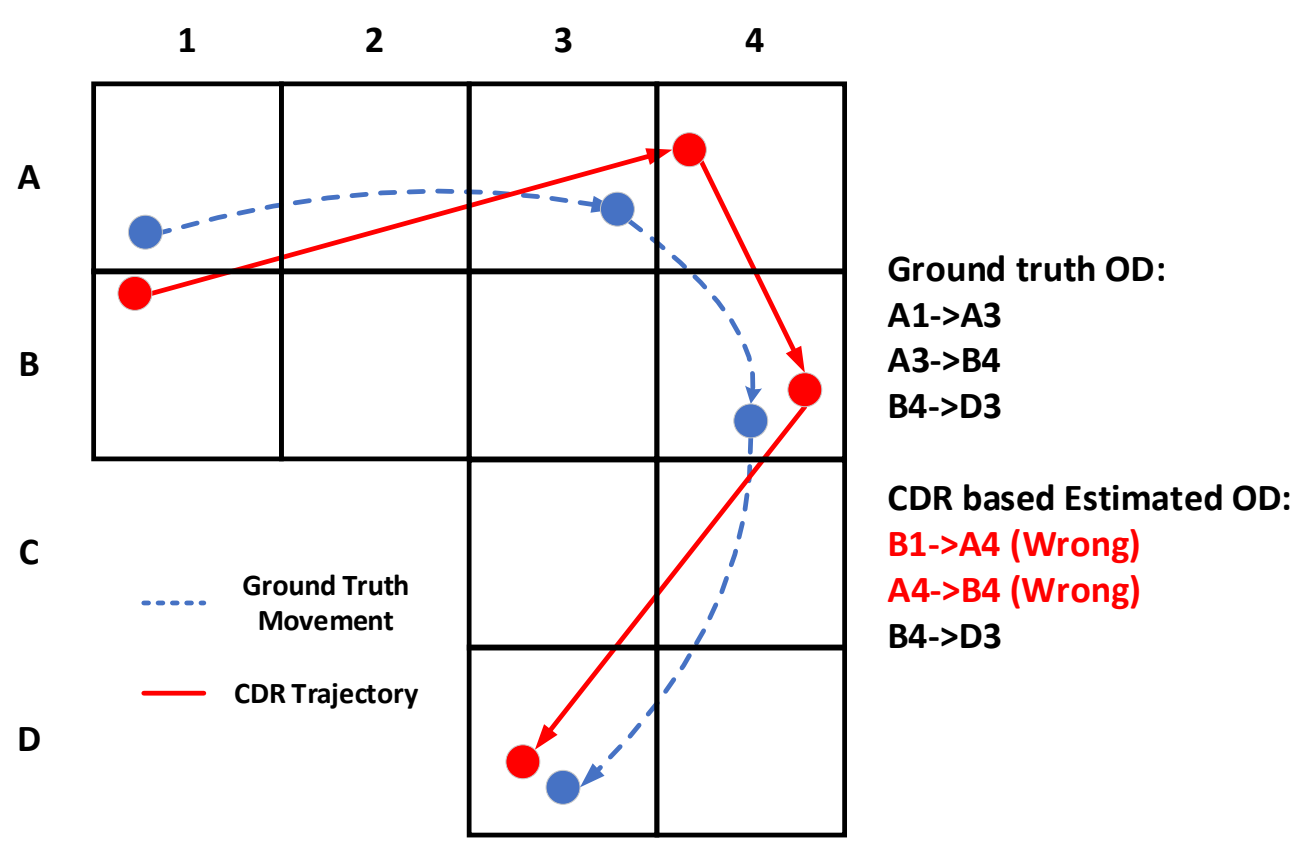

Fig. 9. Comparison between estimated OD pair and ground truth

TABLE 4
SIMILARITY BETWEEN GPS BASED MOBILITY TABLEAU AND CDR BASED MOBILITY TABLEAU

| | 8 /day (*) | 12 /day (*) | 24 /day | 96 /day | 144 /day |
|---|---|---|---|---|---|
| R-squared | 0.57 | 0.59 | 0.31 | 0.68 | 0.82 |
| RMSE | 0.0981 | 0.0951 | 0.1001 | 0.0692 | 0.0557 |
| NMD (Volume Similarity) | 92.11% | 87.67% | 70.40% | 93.96% | 97.99% |
| SP (Spatial Similarity) | 75.19% | 43.77% | 2.32% | 16.74% | 71.43% |
| NMA (Mass Inclusiveness) | 95.46% | 96.97% | 98.21% | 98.66% | 98.81% |
| RRNSA (Structural Similarity) | 84.61% | 82.29% | 96.47% | 94.48% | 91.54% |

∗ : Stay point detection method is different

month.

The density of cell towers of the sampling condition is about 25 cell towers per km² in average, and the CDR records are sampled every 5 minutes corresponding to GPS records. To simulate different sampling frequency in real world, we downsample the data to get different CDR dataset with different number of records per day. When the number of records equals to or be larger than 24/day, we extract OD trip by detecting stay point. When the number is smaller, since CDR is too sparse to detect stay point, we simply assume each CDR record represents a stay point unless the position of the individual did not change compared with the last record.

The result of the comparison is listed in Table 4, which contains SP, NMD, NMA RRNSA in our methods and traditional $R^2$ and RMSE. Compared with 144 records per day, when the number of records reduces, the loss of trips information will cause volume similarity reduces too. When the number of records gets lower than 24 per day, the changing of stay point detection method causes the number of trips grow up to form a rebound. From the table we can conclude findings as following:

- The changing tendency with number of records per day for NMD is similar with $R^2$ and RMSE because they all represent the volume similarity of mobility tableaus. However, RMSE is disturbed by long-distance OD pair that are all zeros, so it is too small to help us judge whether the result is good or not. $R^2$ cannot reflect the geographical correlation, so the value shows the correlation is not high, but our NMD indicates that the volume similarity is considerable indeed.
- When NMD is high, SP is also high, which means the major of differences are spatial error. When NMD is low at 24 records per day, SP is close to zero because at this time differences are caused by information loss instead of sptial error.
- NMA is always close to 100% for 24, 96, 144 records per day, because the information loss will not cause exclusiveness. For 8 and 12 records per day, since the stay point detection is not accurate, NMA slightly gets lower.
- RRNSA is quite different from volume similarity. Although volume similarity is lowest at 24 records per day, structural similarity is the highest because less records bring less spatial error from cell tower, but the general structure is the same. For 8 and 12 records per day, though volume similarity is not low, RRNSA is much lower than 24, 96 and 144 per day because inaccurate stay points break the original structure.

The above findings demonstrate that our metrics have the ability to illustrate the difference between mobility tableaus in high dimension which traditional methods can not provide.

### 7.2 Different City Moblity Tableau Comparison

In this section, we demonstrate the example of finding best match of city slices based on mobility tableau similarity between two different cities to. Tokyo, Japan is selected as source city and Osaka, Japan is selected as target city.

We divide Tokyo into a $40 \times 40$ grid and Osaka into a $15 \times 15$ grid, the cell of which is with about 1km width. The size of slice is selected as $5km \times 5km$, so there are $36 \times 36$ and $11 \times 11$ different slices in Tokyo and Osaka, respectively. mobility tableaus are generated by 1,000 individuals' 1-month GPS data in Tokyo and Osaka.

The heat map of total population flow in different slice in Tokyo and Osaka is shown as Fig. 10. We pick the slice with maximum total population flow in Tokyo as source slice for example to find the most similar slice in Osaka. This slice is around Shinjuku, which is the most popular region in Tokyo. In order to achieve better matching, we consider about different transformations of the region including the combination of rotation and reverse, totally 8 types.

Since the population scale is different in two cities and we are more interested in mobility structure, here we select RRNSA in our method as target metric and compare it with $R^2$ and RMSE.

The heat map of RRNSA, $R^2$ and RMSE for slices with most proper transformation in Osaka is shown as Fig. 11. From the figure we can see that many slices have high $R^2$ with target slice and we cannot find a best match, because in this case slice is not large and all cells in the slice share a common feature that they have population flow with neighbor cells but no flow with far cells. The pattern of RMSE is clearer than $R^2$, but it is still disturbed by the volume of population. Finally, our RRNSA points out the unique best matching slice, which also fits $R^2$ and RMSE.

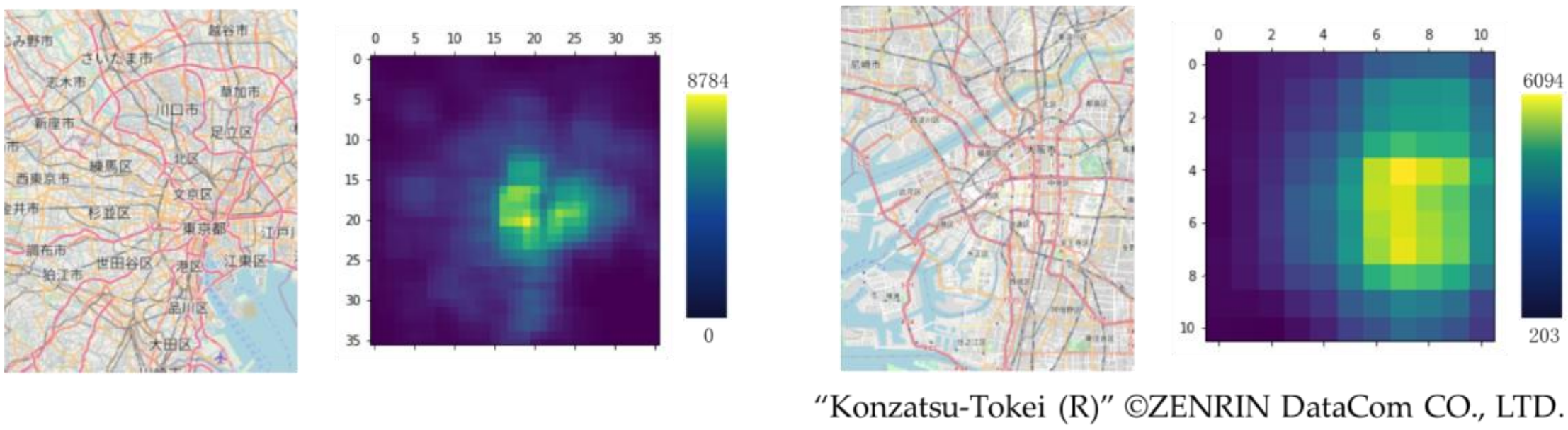

(a) Tokyo Case (b) Osaka Case

Fig. 10. Map and heat map of population flow for each slice in Tokyo and Osaka

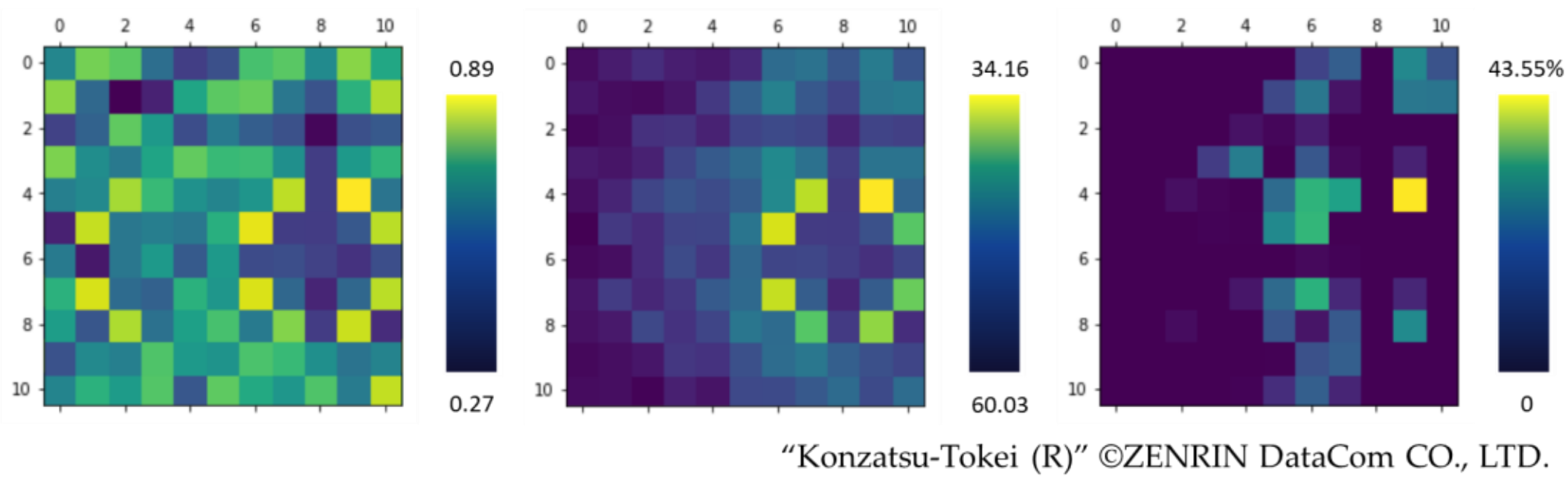

(a) The heat map of $R^2$ (b) The heat map of RMSE (c) The heat map of RRNSA

Fig. 11. The heat map of RRNSA, $R^2$, RMSE for slices with most proper transformation in Osaka

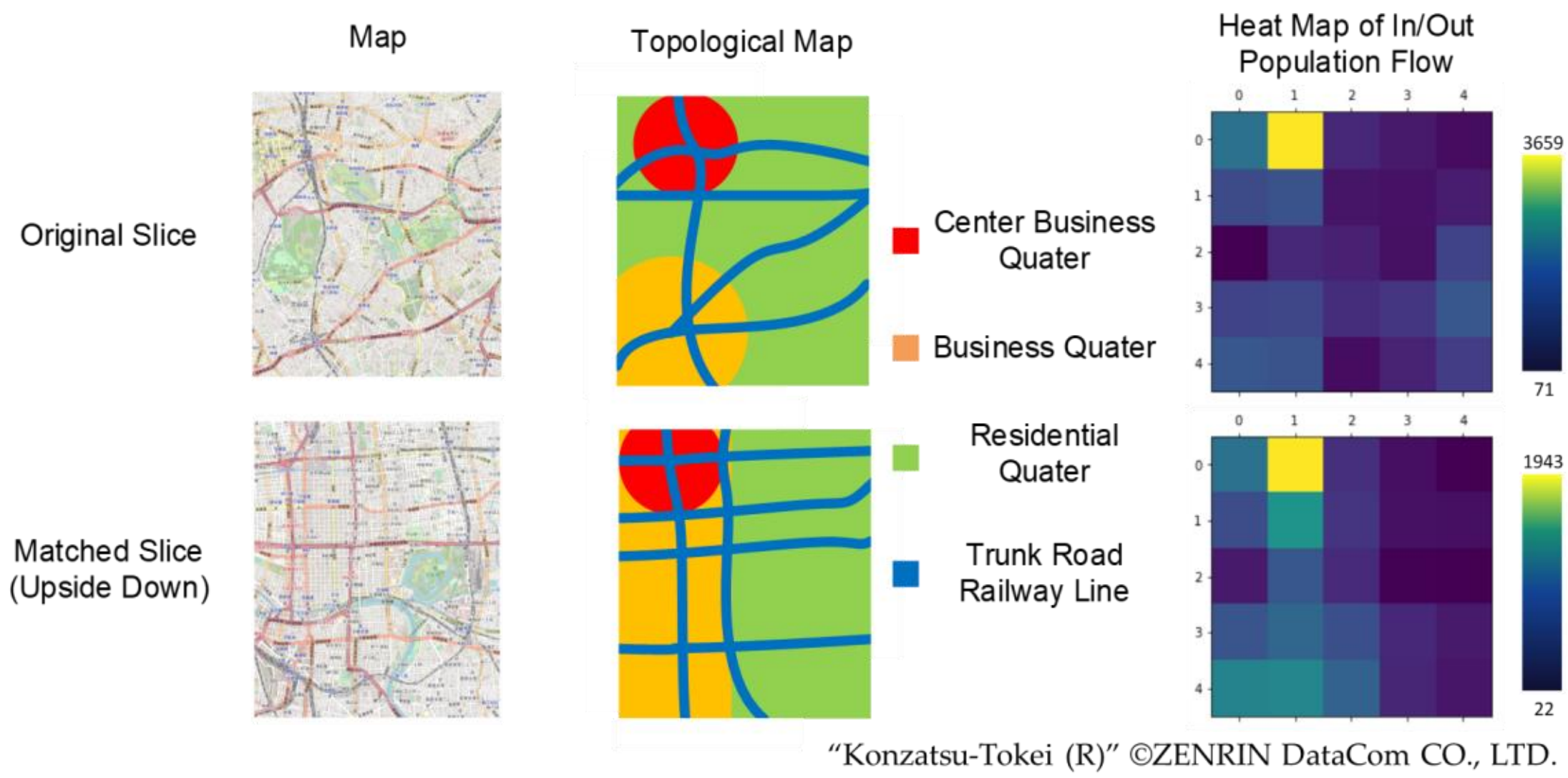

Fig. 12. Comparison between slice in Tokyo and best matching slice in Osaka

To further validate whether two slices are similar in structure, we visualize maps, topological maps and heat maps of in/out population flow as shown in Fig. 12. From the topological maps we can see that two slices have very similar structure. In original slice of Tokyo, the left part is two business quarters: Shinjuku and Shibuya, while in the matched slice (upside down) in Osaka, the left part is also a business quarter in Chuoku, which is the most popular area in Osaka. The geographical positions of two center business quarters in slices are also very similar. Hence, the right parts of two slices are both far from popular center, so they are both relatively lack of population flow. The heat map of in/out population flow validates these features, since in both two slices there is just one cell at the same position prominently full of population flow and the left parts of two slices are both much more popular than right part. Therefore, we can conclude that the matching by our metrics is reasonable.

## 7 Conclusion

By expanding origin-destination matrix, a novel concept named mobility tableau is proposed in this paper, which is a geographic correlation related aggregated tableau representation to the population flow distributed between different location pairs of a study site. The similarity between two mobility tableaus provides several robust statistical metrics to explain significant physical meanings, including mass similarity, spatial similarity, mass inclusiveness, structural similarity, etc. An innovative mobility tableau similarity measurement method is proposed by optimizing the least transforming Manhattan distance to transform the vector graph of one mobility tableau into the other and is optimized to be efficient.

A set of sensitive analysis are conducted to validate the robustness of metrics. Two real world application case studies including an estimated mobility tableaus validation and a different cities' mobility tableaus comparison also indicates the method is a holistic measure.

Compared with traditional mobility tableaus comparison methods, the proposed method not only include both deviations of individual population flows and the structural (dis)similarity between the mobility tableaus with geographic correlation, but also provides several robust statistical metrics to explain different significant physical meanings without any parameters.

While the proposed vector map optimization mobility tableaus can successfully measure the similarity between two mobility datasets in the form of mobility tableaus, a major limitation still remains. The issue is that even though the point set oriented Kuhn-Munkres algorithm dramatically reduce the computation time, in some extreme cases the computation time will still achieve $O(n^3)$ scale, where $n$ equals to the total number of regions. In Tokyo case, when we divide Tokyo into $40 \times 40$ grid, $n$ will be 1,600 and the computation time cannot be ignored.

One of the main future research direction is to further improve the point set oriented Kuhn-Munkres algorithm. Since the bipartite graph generated from mobility tableaus transforming process is quite special because each vertex in graph has correlation with other vertexes, there should be some efficient ways to simplify the augment path searching. Then large-scale computation for Manhattan distance between any matrices could be conducted readily. Another future research direction is to expand spatial vector graph transforming into spatial-temporal vector graph transforming. Currently for spatial-temporal mobility tableaus, the similarity needs to be measured independently by dividing mobility tableaus into different time slices. If we expand vector map from 2-dimensional into 3-dimensional and define a proper spatial-temporal cost transforming rule, a more reasonable spatial-temporal mobility similarity measurement can be achieved.